\pdfoutput=1
\documentclass[aps,prl,twocolumn,showpacs,superscriptaddress,a4paper,
    floatfix]{revtex4}
\usepackage[latin1]{inputenc}
\usepackage{graphicx}
\usepackage{amsmath,amssymb}
\usepackage{hyperref}
\usepackage{datetime}
\usepackage{subfigure}
\usepackage{enumerate}

\begin{document}

\title{A thermoelectric heat engine with ultracold atoms}

\author{Jean-Philippe Brantut}
\affiliation{Department of Physics, ETH Zurich, 8093 Zurich, Switzerland}
\author{Charles Grenier}
\affiliation{Centre de Physique Th\'eorique, \'Ecole Polytechnique, CNRS, 
91128 Palaiseau Cedex, France}
\author{Jakob Meineke}
\affiliation{Department of Physics, ETH Zurich, 8093 Zurich, Switzerland}
\author{David Stadler}
\affiliation{Department of Physics, ETH Zurich, 8093 Zurich, Switzerland}
\author{Sebastian Krinner}
\affiliation{Department of Physics, ETH Zurich, 8093 Zurich, Switzerland}
\author{Corinna Kollath}
\affiliation{HISKP, Universit\"at Bonn, Nussallee 14-16, D-53115 Bonn, 
Germany}
\author{Tilman Esslinger}
\affiliation{Department of Physics, ETH Zurich, 8093 Zurich, Switzerland}
\author{Antoine Georges}
\affiliation{Centre de Physique Th\'eorique, \'Ecole Polytechnique, CNRS, 
91128 Palaiseau Cedex, France}
\affiliation{Coll\`ege de France, 11 place Marcelin Berthelot, 75005 
Paris, France}
\affiliation{DPMC-MaNEP, Universit\'e de Gen\`eve, CH-1211 Gen\`eve, 
Switzerland}

\date{\pdfdate}

\begin{abstract}

Thermoelectric effects, such as the generation of a particle current by a 
temperature gradient, have their origin in a reversible coupling between 
heat and particle flows. These effects are fundamental probes for materials 
and have applications to cooling and power generation. Here we demonstrate 
thermoelectricity in a fermionic cold atoms channel, ballistic or 
diffusive, connected to two reservoirs. We show that the magnitude of the 
effect and the efficiency of energy conversion can be optimized by 
controlling the geometry or disorder strength. Our observations are in 
quantitative agreement with a theoretical model based on the 
Landauer-B\"uttiker formalism. Our device provides a controllable 
model-system to explore mechanisms of energy conversion and realizes a 
cold atom based heat engine.

\end{abstract}

\maketitle

In general, heat and particle transport are coupled processes \cite{Onsager:1931aa}. This coupling leads to thermoelectric effects, such as a Seebeck voltage drop in a conductor subject to a thermal gradient. These effects are important for probing elementary excitations in materials, for example, giving access to the sign of charge carriers \cite{ashcroft2002physique}. Moreover, they have practical applications to refrigeration, and power generation from waste-heat recovery~\cite{Goldsmid,snyder2008com}. Recently, there has also been interest in thermoelectric effects in nano- and molecular-scale electronic devices \cite{entin_2010,sanchez_2011}. The progress in modeling solid-state physics with cold atoms \cite{Esslinger:2010aa,Bloch:2012aa} raises the question whether thermoelectricity can be observed in such a controlled setting~\cite{Grenier:2012tg,Kim:2012aa}, where set-ups analogous to mesoscopic devices were realized \cite{Albiez:2005aa,Ramanathan:2011aa,Brantut:2012aa}. Whilst the thermodynamic interplay between thermal and density collective modes has been seen in a second sound experiment \cite{Sidorenkov:2013aa}, thermoelectric effects have so far not been investigated.

Here, we demonstrate a cold atoms device in which a particle current is 
generated by a temperature bias. 
We prepare a mesoscopic channel connecting two atomic 
reservoirs having equal particle numbers. Heating one of the 
reservoirs establishes a temperature bias and the compressible cloud 
forming the hot reservoir expands. Hence, one naively expects an initial 
particle flow from the cold denser side to the hot. In contrast, we 
observe the opposite effect: a net particle current initially directed from 
the hot to the cold side. This is a direct manifestation of the 
intrinsic thermoelectric power of the channel. The temperature bias leads 
to a current of high-energy particles from hot to cold and a current of low-energy particles from cold to hot. In our channel, particles are transported at a 
rate which increases with energy, leading to an asymmetry between the high-
energy and low-energy particle currents. This results in a total current 
from hot to cold, which overcomes the thermodynamic effect of the 
reservoirs. Hence, work is performed by carrying atoms from lower to higher 
chemical potential, and our system can be regarded as a cold-atoms based 
heat engine.

\begin{figure}[htb]
    \includegraphics{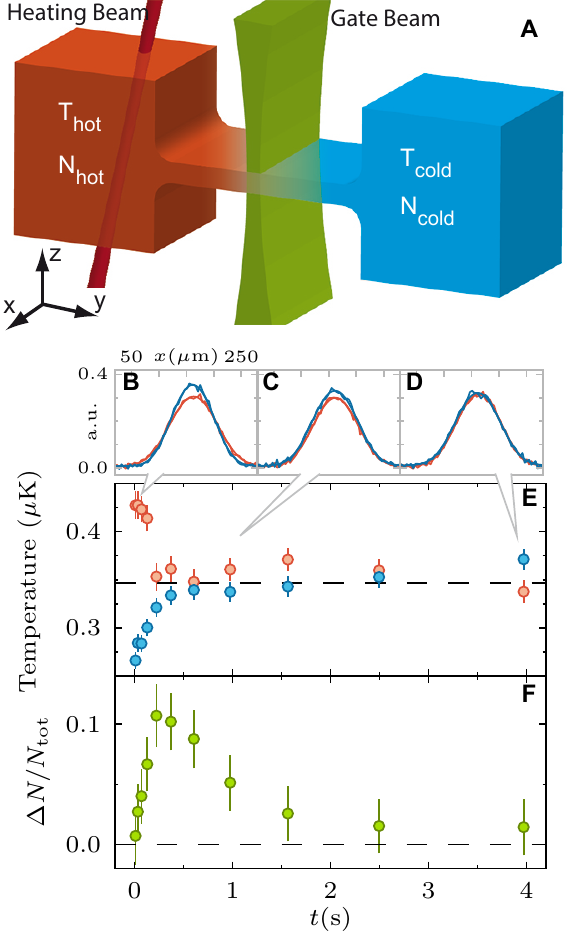}
    \caption{
    Concept of the experiment. A: A quasi-two dimensional channel connects 
two atomic reservoirs. A gate beam intersects with 
the channel and blocks particle and heat transport. A heating beam 
traverses the left reservoir and heats it in a controlled way. 
B-D: Line-sums along $y$ of the density in the hot (red) and cold (blue) 
reservoir at different evolution times (0.01, 0.98 and 3.97\,s) after time-of-flight. E: $T_h$ (red) and $T_c$ (blue) 
as a function of time. Dashed line~: $\bar{T}$ at the initial time. F: $\Delta N/N_{tot}$ as a function of time. In the channel 
$\nu_z$ was set to $3.5\,$kHz with a disorder of average strength 
$542\,$nK~(see text).
}
    \label{fig:concept}
\end{figure}

A schematic view of the experimental setup is shown in figure 
\ref{fig:concept}A. It is based on our previous work on conduction 
processes in Fermi gases \cite{Brantut:2012aa,Stadler:2012kx}. Initially we prepare $N_{\mathrm{tot}}=3.1(4)\cdot10^5$ weakly interacting $^6$Li atoms at a temperature of 
$250(9)\,$nK in an elongated trap, where the Fermi temperature of the cloud 
is $T_F=931(44)\,$nK. Using a 
repulsive laser beam (not shown) having a nodal line at its center, the cloud is then separated into two identical reservoirs connected by a channel. Tuning the power of the beam allows to adjust the trap frequency $\nu_z$ in the channel up to 
$10$\,kHz. We then raise a gate potential in the channel, preventing any 
energy or particle exchange between the reservoirs. A controlled heating 
of the left reservoir is performed, increasing its temperature by 
typically $200$\,nK. Afterwards, the gate potential is removed abruptly and the system evolves for a variable 
time. For the observations, the gate potential is raised again, the power 
of the laser beam creating the channel is ramped to zero in $400$\,ms, and 
each reservoir is left to equilibrate independently for $100$\,ms 
\cite{materialsandmethods}.

We take absorption images of the cloud and measure the temperatures $T_h$ ($T_c$) and atom number $N_h$ ($N_c$) in the hot (cold) reservoir 
using finite temperature Fermi fits, as indicated in figure 
\ref{fig:concept}B-D.
This allows us to reconstruct the time evolution of the number
imbalance $\Delta N = N_c - N_h$ and temperature bias $\Delta T = T_c - 
T_h$. Figure\ref{fig:concept} E and F show typical results. The 
temperatures equilibrate fast,
similar to the equilibration of atom numbers observed in the case of pure 
atomic flow \cite{Brantut:2012aa}. In contrast, the atom-number difference in figure 
\ref{fig:concept}F starts at zero, first grows 
fast, and then decreases back to zero. This transient atomic current is the 
fingerprint of thermoelectricity.

To explain quantitatively our observations, we model the channel as a 
linear circuit element having conductance $G$, 
thermal conductance $G_T$, and
thermopower $\alpha_{ch}$. In this framework, the particle and entropy 
currents flowing in the channel are given by~:
\begin{equation}\label{eq:lin_response}
\left(
\begin{array}{c}
 I_N\\
 I_S
\end{array}
\right)
=
-G\left(\begin{array}{cc}
		1&\alpha_{ch}\\
	      \alpha_{ch}&L+\alpha_{ch}^2
             \end{array}
\right)
\left(
\begin{array}{c}
 \mu_c-\mu_h\\
 T_c-T_h
\end{array}
\right)\,.
\end{equation}
In this expression, $I_N = \partial\Delta N/\partial t$ and $I_S = 
 \partial \Delta S/\partial t$, with $\Delta S = S_c-S_h$.
Further $\mu_h$, $\mu_c$ are the chemical potentials of the hot and cold 
reservoirs, $L=G_T/\bar{T}G$ is the Lorenz 
number~\cite{ashcroft2002physique} of the 
channel, and $\bar{T}=(T_h + T_c)/2$. Combining 
eq.~\eqref{eq:lin_response} 
with the thermodynamics of the reservoirs leads to the equation 
for the time evolution of $\Delta N$ and $\Delta T$~:
\begin{equation}\label{eq:temp_imb}
 \tau_0
\frac{d}{dt}\left(
\begin{array}{c}
 \Delta N\\
 \Delta T
\end{array}
\right)
=
-\left(\begin{array}{cc}
                                 1 &-\kappa(\alpha_r-\alpha_{ch})\\
                                 -\frac{\alpha_r-\alpha_{ch}}{\ell\kappa} & 
\frac{L+(\alpha_r-\alpha_{ch})^2}{\ell}
                                \end{array}\right)
\left(
\begin{array}{c}
 \Delta N\\
 \Delta T
\end{array}
\right)\,.
\end{equation}
Here, $\kappa=\left. \frac{\partial N}{\partial 
\mu}\right|_{T}$, $C_N=\left. T\,\frac{\partial S}{\partial 
T}\right|_{N}$ and $\alpha_r = \left. \frac{\partial S}{\partial 
N}\right|_{T}$ are the compressibility, specific heat and 
dilatation coefficient of each reservoir, calculated at the average 
temperature $\bar{T}$ and particle number $(N_c+N_h)/2$. $\ell 
= \frac{C_N}{\kappa \bar{T}}$ is an analogue of the Lorenz number for the 
reservoirs, and $\tau_0 = \kappa G^{-1}$ is the particle transport 
timescale, analogous to a capacitor's discharge 
time~\cite{Brantut:2012aa}. 
Eq.~\eqref{eq:temp_imb} shows that the thermoelectric response results from 
the competition between the entropy transported through the channel, 
described by $\alpha_{ch}$, and the entropy created by removing one atom 
from one reservoir and adding it to the other, described by $\alpha_r$. 
The channel properties $G,G_T$ and $\alpha_{ch}$ are 
calculated using the Landauer-B\"{u}ttiker 
formalism~\cite{RevModPhys.71.S306,
Buttiker:1988:SEC:49387.49388,materialsandmethods}. 

\begin{figure}[htb]
    \includegraphics{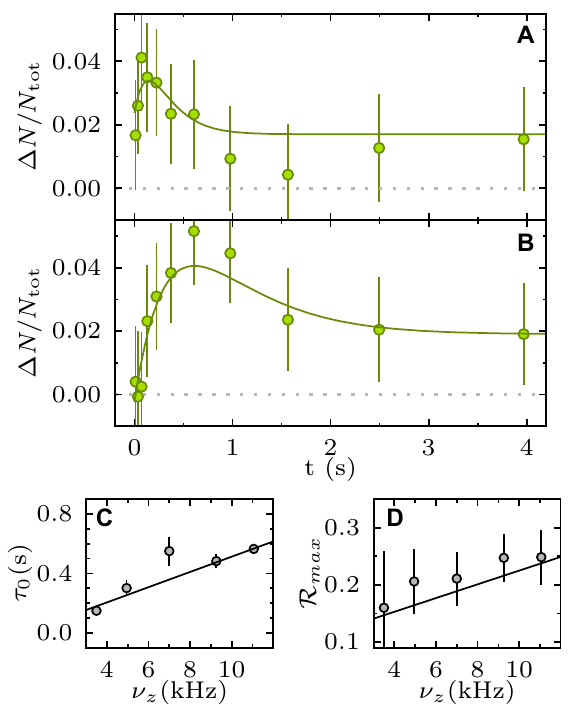}
    \caption{
    Thermoelectric response of a ballistic channel for various confinements. 
    A and B: Time evolution of $\Delta N/N_{tot}$ for $\nu_z = 3.5\,$kHz 
(A) and $\nu_z = 9.3\,$kHz (B), compared to theory (solid line, see 
text and \cite{materialsandmethods}).
     C and D: Timescale $\tau_0$ and maximum thermoelectric response as a function of confinement. Symbols are the fitted values from experiment 
and the solid line is the theoretical prediction. 
}
    \label{fig:ballistic}
\end{figure}

The geometry of the channel influences its thermoelectric 
properties~\cite{Heremansdresselhaus}. We 
measured the transient imbalance and the temperature evolution for
various confinements in a ballistic channel. Two examples for $\nu_z=3.5$ and $9.3$\,kHz are presented in figure 
\ref{fig:ballistic}A and B. The evolution of both 
$\Delta N$ and $\Delta T$ is faithfully described by the theoretical 
model, which 
does not involve any adjustable 
parameter~\cite{materialsandmethods}. As shown in figure 
\ref{fig:ballistic}C, the dynamics of both temperature
and atom number evolution is slowing down with increasing $\nu_z$, as 
expected since the 
number of conducting modes is reduced. 
The experimental values are extracted from fitting the experimental data with the 
theoretical model with $\tau_0$ as the only free 
parameter~\cite{materialsandmethods}. They agree well with the theoretical 
predictions and independent measurements performed 
with a pure particle number imbalance~\cite{materialsandmethods}.

At the same time the amplitude of the transient imbalance increases with 
stronger confinement. This can be qualitatively 
understood by noting that the energy dependence of the density of states in 
the channel is enhanced with increasing trap frequencies (see 
\cite{materialsandmethods}). We define the thermoelectric response 
$\mathcal{R} = (\Delta N /N_{\mathrm{tot}}) / 
(\Delta T_0/T_F)$ where $\Delta T_0$ is the initial temperature difference $\Delta T_0$~\cite{materialsandmethods}.
In figure \ref{fig:ballistic}D we present the maximum thermoelectric 
response $\mathcal{R}_{\mathrm{max}}$ which displays an approximately linear 
increase with $\nu_z$.

\begin{figure}[htb]
    \includegraphics{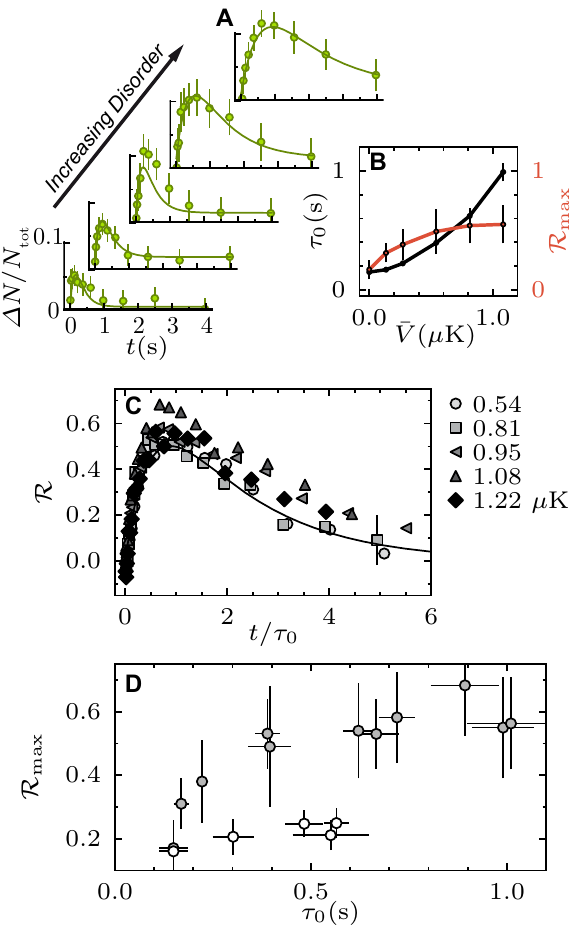}
    \caption{
    Thermoelectric response in the ballistic to diffusive
crossover. 
A: Time evolution of $\Delta N/N_{tot}$ for increasing disorder
strength, for a fixed confinement of $\nu_z = 3.5\,$kHz.
Solid lines: theory. 
B: Fitted timescale $\tau_0$ (black) and $\mathcal{R}_{max}$ 
(red) as a function of disorder strength for the data set shown in A. 
C: Rescaled time evolution of $\mathcal{R}$ (see 
text) in the regime of strong disorder from $\bar{V} = 542\,$nK to 
$1220\,$nK and 
fixed $\nu_z = 4.95\,$kHz. Black line: theoretical calculation. 
Here, $\mathcal{R}$ depends only on the overall timescale, 
and all the curves collapse after rescaling the time axis.
D: Comparing $\mathcal{R}_{\mathrm{max}}$ as a function of 
timescale for the diffusive (gray points) and ballistic case (open 
circles).
}
    \label{fig:diffusive}
\end{figure}

We now investigate the effects of disorder on the thermoelectric properties of  
the channel. We project a random potential of adjustable strength 
$\bar{V}$ onto the channel in the form of a blue detuned laser speckle 
pattern \cite{materialsandmethods}. Figure \ref{fig:diffusive}A presents 
the time evolution of $\Delta N/N_{tot}$ for increasing disorder, for 
fixed $\nu_z = 3.5\,$kHz. First we observe that the time scale $\tau_0$ of the transport process increases: 
the resistance increases as the channel crosses over from ballistic to 
diffusive. In addition to this slowdown, $\mathcal{R}_{\mathrm{max}}$ increases from $0.17(8)$ without disorder to $0.55(16)$ 
for a strong disorder of $1.1\,\mu$K. 

For the strongest disorder, we observe (Fig.~\ref{fig:diffusive}B)
that the thermoelectric response saturates, while the timescale $\tau_0$ keeps increasing, 
indicating the continuous increase of resistance with disorder. 
To further investigate this point, we performed experiments for a fixed confinement in the channel of $4.95\,$kHz, 
and several large disorder strengths ranging from $542$ to $1220$\,nK.
As illustrated in Fig.~\ref{fig:diffusive}C, we find that, in this 
regime, the full data-set collapses to a single curve, provided the time axis is rescaled by the 
timescale $\tau_0$ extracted from an independent atomic conduction 
experiment \cite{Brantut:2012aa,materialsandmethods}. 
This shows that thermoelectricity is independent of the actual resistance, 
as expected from the fact that thermopower is a ratio of linear response 
coefficients. This makes thermopower less sensitive than resistance to the 
details of the conductor, a fact widely 
used in condensed matter physics \cite{Mokashi:2012aa}.

The effect of disorder can be described by extending our theoretical model \cite{materialsandmethods}, 
introducing an energy-dependent transparency of the constriction. This 
transparency involves the energy-dependent 
mean-free path in the channel, product of the particle velocity and scattering time. 
At strong disorder, the scaling of the scattering time with disorder is modelled assuming that the energy-dependence of the scattering time can be neglected. 
When expressed as a function of $t/\tau_0$, a unique theoretical scaling curve independent of $\bar{V}$ is predicted for the thermoelectric response, which 
describes well the experimental data (Fig.~\ref{fig:diffusive}C). 
The time-scale $\tau_0$ is the only adjustable parameter in this strong disorder regime. A fit to the 
experimental data allows one to extract the dependence of $\tau_0$ 
on disorder strength \cite{materialsandmethods}. 

Extrapolating this dependence into the weak disorder regime,
 the resulting theoretical curves for the transient particle imbalance 
accurately describe the data over the entire range of disorder strengths 
(Fig.\ref{fig:diffusive}A).

Confinement and disorder are two independent ways of influencing the thermoelectric properties. 
To compare their respective merits, we display $\mathcal{R}_{\mathrm{max}}$ as a 
function of $\tau_0$ in Fig~\ref{fig:diffusive}D. This shows that disorder 
is more efficient than confinement to increase thermoelectricity. For 
the largest time scales, we observe a $\sim 3$-fold increase in the 
diffusive case compared to the ballistic one. This is due to the stronger 
energy-dependence of transmission in the diffusive compared to 
the ballistic case.

\begin{figure}[htb]
    \includegraphics{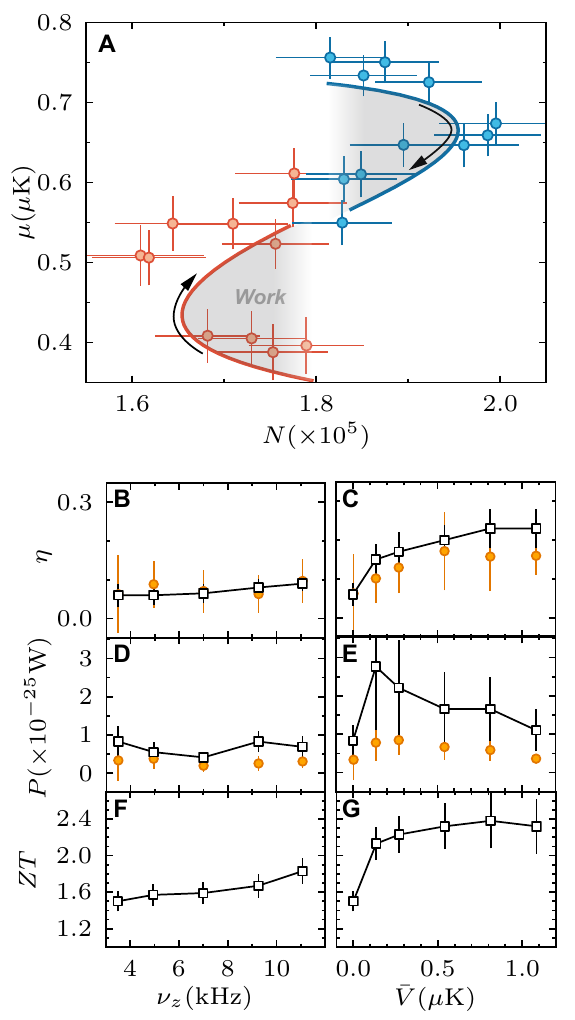}
    \caption{
    Our system as a heat engine. 
     A: Evolution of the hot (red) and cold (blue) reservoir in the 
$\mu$-$N$plane for $\nu_z = 3.5\,\mathrm{kHz}$ and 
$\bar{V}=542\,\mathrm{nK}$. 
     Experiments: symbols, Theory: solid lines. The black arrows indicate 
the direction of time and the sum of the enclosed areas yields the total 
work. B, D, F: Efficiency, power and dimensionless figure of merit of 
the channel in the ballistic case, as a function of confinement. 
     C, E, G: The same quantities as a function of disorder strength for $\nu_z = 3.5\,$kHz. 
     Orange symbols: experiments; Black symbols: theory (theoretical error 
bars estimated from the uncertainties on the input parameters).
}
    \label{fig:efficiency}
\end{figure}

In the experiment, a controlled exchange of heat between a hot and a cold 
reservoir is used to produce a directed current, i.e. work. This motivates 
an analysis in terms of heat engines. To do so, we evaluate the 
work, efficiency, and power of the process. The 
area enclosed in the $\mu-N$ plane 
(Fig.~\ref{fig:efficiency}A)
 represents the work $W = 1/2\int_0^\infty dt 
(\mu_c-\mu_h)(t) I_N(t)$ produced during the evolution, which we evaluate from $N_{c,h}$ and 
$T_{c,h}$. Similarly, we evaluate the 
heat associated to the transport process $Q = -1/2\int_0^\infty dt 
(T_c-T_h)(t) I_S(t)$  \cite{materialsandmethods}.

We then introduce the relative efficiency 
$\eta=W/Q$ ~\cite{Goldsmid,materialsandmethods}. For a 
reversible process $\eta=1$. We find that $\eta$ is largest for 
configurations 
where the thermoelectric response is largest (Fig~\ref{fig:efficiency}B 
and C). The large value of the 
efficiency suggests that the channel is a very good thermoelectric 
material. Generally, the efficiency increases as the dynamics slows down, 
since the thermodynamic processes become closer to reversibility. 
Therefore, a complementary criteria to evaluate the merits of the various 
thermoelectric configurations is the 
cycle-averaged power~\cite{Callen:Thermodynamics}, 
estimated by $W/\tau_0$.  As shown in figure \ref{fig:efficiency}D and E, 
the power goes down for the largest resistances, since the increase in 
work is overcompensated by the slow-down of the process.

We now focus on the channel, independently of 
the reservoirs. We use the transport coefficients extracted using our model 
to estimate the dimensionless figure of merit $ZT= \alpha_{ch}^2  /L$ 
\cite{Goldsmid} of 
the channel (Fig~\ref{fig:efficiency}F and 
G). This number is related to 
the efficiency achievable with a channel operating at maximum 
power, and is used as a criterion for engineering thermoelectric devices 
\cite{DiSalvo:1999aa}. For the largest thermoelectric response 
observed, we infer $ZT=2.4$, which is among the largest values observed in 
any solid-state material~\cite{snyder2008com}.

Our experiment 
demonstrates thermoelectric effects in quantum gases and shows that 
thermopower is a sensitive observable in this context. 
We have used it as a probe for the ballistic or diffusive nature of 
transport, independently from resistance. A high 
efficiency and figure of merit compared to many materials are found due to 
the absence of phonon contributions to the heat 
conductivity~\cite{Goldsmid}. Our technique 
can be straightforwardly generalised to interacting systems, where 
thermoelectric properties are of fundamental interest 
\cite{Behnia:2004aa,Zhang:2011aa,Micklitz:2012aa}. 
The reversed operation of our device leads in principle to cooling by the Peltier 
effect. This may be useful to cool quantum gases to low entropy, needed to 
explore strongly correlated fermions in lattices.

The Z\"urich team acknowledges fruitful discussions with J. Blatter at the 
initial stage of the experiment and CK thanks M. B\"uttiker for helpful 
discussions. We acknowledge financing from NCCR MaNEP and QSIT of the SNF, 
the ERC Project SQMS, the FP7 Project NAME-QUAM, the ETHZ Schr\"odinger 
chair, the DARPA-OLE program, ANR (FAMOUS), and ETHZ. J.P.B. is partially 
supported by the EU through a Marie Curie Fellowship.

%

\renewcommand*{\citenumfont}[1]{S#1}
\renewcommand*{\bibnumfmt}[1]{[S#1]}

\onecolumngrid
\pagebreak
\appendix

\section*{Materials and Methods}

\subsection*{Details on the experimental setup}

\subsubsection*{Preparation of the clouds}

The preparation of the clouds follows the method used in our previous studies \cite{Brantut} with minor modifications. The $^6$Li atoms are prepared in an incoherent mixture of the lowest and third hyperfine states, and placed in a homogeneous magnetic field of 388\,G, 
where the scattering length is about $-800$\,a$_0$. The cloud is then
evaporatively cooled in a hybrid magnetic and optical dipole trap at 
$1064$\,nm, with a waist of $21\,\mu$m as in \cite{Brantut}. To avoid anharmonicity and thus allow accurate thermometry, the 
cloud is then transferred in a dipole trap propagating along the same 
direction with a waist of $70\,\mu$m. We use a final laser power of 
$800$\,mW for this trap, yielding transverse trap frequencies of $414\,$Hz, 
with negligible ellipticity. At this laser power, the trap depth is $6.3\mu$K, ensuring that no further evaporation takes place during the measurements.

After the transfer into the final dipole trap, the magnetic field is increased up to 552\,G, where the 
scattering length is about $-100$\,a$_0$. At this magnetic field, the longitudinal trap frequency originating from the curvature of the field is 
$23.5\,$Hz, measured using the dipole oscillations of a cloud.

\subsubsection*{Trap configuration}

Like in \cite{Brantut}, we superimpose on the cloud a laser beam at $532$\,nm having a TEM$_{01}$-like mode profile \cite{Meyrath:2005aa}, propagating along the $x$ direction. This creates the channel at the center of the trap, connected on both sides to the reservoirs. The waist of this laser beam along 
the long direction of the cloud ($y$ direction) is $30\,\mu$m, about ten times shorter than the cloud. Varying the power of this laser beam allows to tune the confinement along the $z$ direction in the channel while leaving the reservoirs unchanged.

We create disorder by illuminating the channel with a speckle pattern created by a laser at $532$\,nm and propagating along the $z$ direction. The speckle pattern is centered on the channel and has an envelope with a $1/e^2$ radius of $36.1$ and 
$40.8\,\mu$m along the $x$ and $y$ directions. The correlation length of the intensity distribution is 
$370\,$nm ($1/e$ radius of the autocorrelation function evaluated using a Gaussian fit). The disorder strength is the 
disorder-averaged potential at the centre of the pattern, evaluated from the total laser power and the polarisability of lithium. 

The gate beam, which is used to isolate the two reservoirs from each other,
is created using a laser beam at $532\,$nm propagating also along the $z$ direction and focused onto the center of the
channel, with waists of $43.3$ and $2.6\,\mu$m along the $x$ and $y$ directions. When turned on, the height of the
potential hill is larger than $30\,\mu$K, thereby ensuring the complete isolation of each reservoir.

The heating beam consists in a laser beam at $767\,$nm propagating along the $z$ direction with a small angle so that it is focused in one of the reservoirs with a waist of $4.95(2)\,\mu$m. Once the cloud has been separated in two independent reservoirs by the channel and the gate beam (see the main text for the description of the sequence), the heating beam is ramped up and sinusoidally modulated at $660$\,Hz for $1$\,s. The laser beam is then switched off before the gate beam is removed.

\subsection*{Systematics and error bars}

\begin{figure}
	\def \thefigure{S1}
	\centering
		\includegraphics[width=0.6\textwidth]{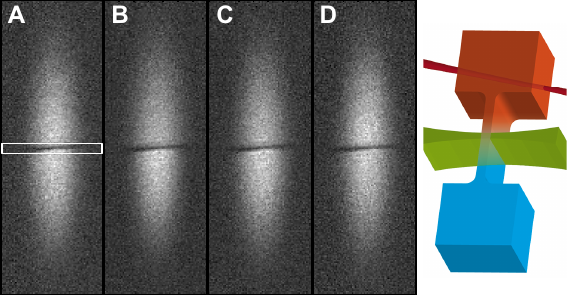}
	\caption{Averaged images of the two reservoirs after the transport process has taken place and after time of flight. Images A,B,C,D are taken for evolution times of $0.01,\,0.37,\,0.98,\,3.97$ s respectively, where A,C,D correspond line sums shown in fig. 1B,C,D in the main text. Image B is taken where $\Delta N/N_{\mathrm{tot}}$ is maximal. The thin region without atoms is the position of the gate beam. The white box in picture A indicates the part that is excluded for the fitting. The schematic view on the right side illustrates the actual configuration of the experiment.}
	\label{fig:fig1_som}
\end{figure}

Temperature, $\Delta N/N_{\mathrm{tot}}$ and their errors are inferred from 2-dimensional Fermi fits performed on typically three averaged pictures (see, for example fig. 1 in the main text and fig. \ref{fig:fig1_som} here). We separately fit the two reservoirs and checked that it gives consistent results with a fit on a full cloud that has twice the atom number of one reservoir. We also checked that the atom number inferred from simply counting and the Fermi fit is the same within our measurement uncertainty. The position of the gate beam defines the separation of the two reservoirs. In order to avoid effects from the gate beam on the Fermi fits we exclude a band of $\pm 4 $ pixels around the center. The center position has to be carefully adjusted. A misalignment by 1 pixel gives a systematic shift of $\Delta N/N_{\mathrm{tot}} \lesssim 0.02$.

Throughout the text, the error bars represent one standard deviation.
The errors on the total atom number and $T_F$ indicated in the main text are the standard deviation evaluated between all the different data sets displayed in the paper. Within a given set, corresponding to one choice of disorder or confinement, the fluctuations of atom number are at most 5\,\%, the temperature fluctuations are below 8\,\%.
The error bars displayed for the time constant $\tau_0$ are fitting errors and for $\mathcal{R}$ they are the combination of the error in $\Delta N/N_{\mathrm{tot}}$ and the initial temperature difference $\Delta T_0$. 

The error bars on the experimentally measured efficiencies and power displayed in fig. 4 in the main text are coming from gaussian error propagation of the involved quantities $\mu, N, T, S, \tau_0$. The error bars on the theoretically expected efficiencies, power and $ZT$ result from the propagation of the uncertainties on the experimentally determined parameters.

\subsection*{Response to initial temperature imbalance}

\begin{figure}[htb]
\def \thefigure{S2}
\includegraphics[width=0.6\textwidth]{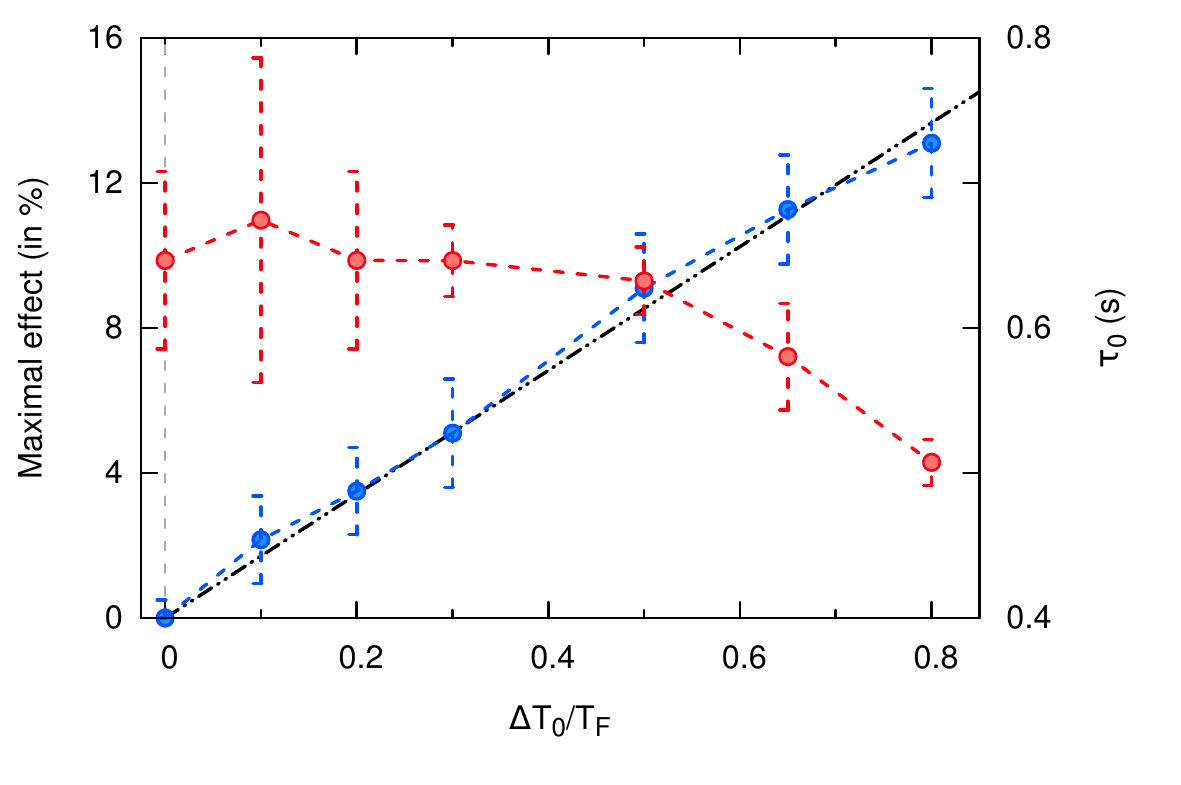}
 \caption{Evolution of the maximal relative imbalance as a function of the 
initial temperature difference, in blue (left axis). The black curve is a 
linear fit. The amplitude of the thermoelectric effect remains linear for 
large values of the initial temperature difference. Timescale vs initial 
temperature difference, in red (right axis).}
\label{fig:linResp_amp_ts}
\end{figure}

We have repeated our measurements on the disordered channel having
$4.95$\,kHz confinement  and speckle power $542$\, nK with various
temperature bias.
 The maximum atom number imbalance observed during the time evolution is
presented on figure \ref{fig:linResp_amp_ts}. The linear increase of the
imbalance with heating supports that our measurements remain in the linear
response regime, despite the relatively large temperature bias. 

In addition to the amplitude, a supplementary check on the timescale has
been performed, and is also reported in fig.~\ref{fig:linResp_amp_ts}. It
shows that the timescale remains constant over a wide range of initial
temperature difference, indicating the linear behavior. Only for the
largest values of $\Delta T_0$ one sees a inclination towards shorter
particle timescales, which we interprete as a sign of a deviation from
linear response. The values of $\frac{\Delta T_0}{T_F}$ considered in the 
main text are between $0.18$ and $0.25$, a range in which we expect our 
linear response model to be valid.

\subsection*{Theoretical model}

The transport setup under consideration is depicted in
Fig.1 in the main text. Two reservoirs of fermionic atoms are connected 
along the $y$-direction by a constriction. The reservoirs are described as 
three-dimensional noninteracting trapped 
Fermi gases, in a harmonic potential in the $x-z$ directions, and 
half-harmonic in the $y$-direction. The reservoir state is characterized 
by 
the temperature $T_{h,c}$ and their chemical potential $\mu_{h,c}$, where 
$h,c$ label the hot and cold reservoir, respectively. The particle number 
$N(T,\mu)$ and entropy $S(T,\mu)$ of each reservoir are given through the 
grand-canonical equation of state. It is convenient to introduce the 
average of a quantity $X$ defined by $\bar{X}=\frac 1 2 (X_h+X_c)$ and its 
difference $\Delta X= (X_c-X_h)$. The channel is oriented along direction 
$y$ and a harmonic confinement is present in its transverse ($x-z$) 
direction. We take the temperature and chemical potential of the channel 
to be the average temperature and chemical potential of the reservoirs. In 
the following we describe in more detail how to obtain the transport 
equation~\eqref{eq:Transport_equation} (eq (2) in the main text) for this 
setup within linear response.

\subsubsection*{Model for the reservoirs}

In the described reservoir configuration, the energy 
levels are given by $\varepsilon =
\hbar\omega_x(n_x+1/2)+\hbar\omega_z(n_z+1/2)+\hbar\omega_y(2n_y+3/2)$, 
where $n_i$ labels the energy level and $\omega_i$ the trapping frequency 
along direction $i$. 
The Fermi energy  of the reservoirs is typically $E_F \simeq 930\,$nK and 
their temperature $T/T_F \simeq 0.25$, such that we are in the limit 
where $E_F, k_BT \gg \text{max}(\omega_x,\omega_y, \omega_z)$.
Hence, one can neglect the discrete structure of the energy levels and the 
thermodynamic properties of the reservoirs, the compressibility $\kappa$, 
the dilatation coefficient $\gamma$ and the heat capacity $C_{N}=\left. 
\frac{\partial S}{\partial T}\right|_{N}$, are well captured by the 
following formulae, with the density of states $g_{r} (\varepsilon)=
\frac{\varepsilon^2}{4(\hbar\omega_x\hbar\omega_y\hbar\omega_z)}$~:
\begin{eqnarray}
\label{eq:thcoeffres_1}
 \kappa =\left. \frac{\partial N}{\partial \mu}\right|_T &=& 
\int_0^\infty 
d\varepsilon
g_{r}(\varepsilon)\left(-\frac{\partial f}{\partial \varepsilon}\right)\\
 \gamma  =\left.\frac{\partial 
N}{\partial T}\right|_\mu=\left.\frac{\partial 
S}{\partial\mu}\right|_T&=& \int_0^\infty d\varepsilon
g_{r}(\varepsilon) \left(\varepsilon-\mu\right)\left(-\frac{\partial 
f}{\partial
\varepsilon}\right)
\label{eq:thcoeffres_2}\\
\label{eq:thcoeffres_3}
 \frac{C_N}{T} + \frac{\gamma^2}{\kappa}  &=& \int_0^\infty d\varepsilon
g_{r}(\varepsilon) \left(\varepsilon-\mu\right)^2\left(-\frac{\partial 
f}{\partial \varepsilon}\right)\,,
\end{eqnarray}
where $f(\varepsilon) = \frac{1}{1+e^{\beta(\epsilon-\mu)}}$ is the 
Fermi-Dirac distribution. The thermodynamic coefficients are calculated 
at the average temperature $\bar{T}$, particle number $(N_c+N_h)/2$ and 
chemical potential $(\mu_c+\mu_h)/2$. In the main text, the dilatation 
properties of the gas are represented by the reservoir contribution 
$\alpha_r = \frac{\gamma}{\kappa}$ to the total thermopower.

\subsubsection*{Model for the channel}
The channel is modeled by a linear circuit element at the 
average temperature $\bar{T}$ and chemical potential $\bar{\mu}$. 
Its linear transport coefficients are given by the following
expressions :
\begin{eqnarray}
\label{eq:conductance}
 G &= \frac{1}{h}\int_0^\infty d\varepsilon\, \Phi(\varepsilon)
\left(-\frac{\partial f}{\partial \varepsilon}\right)\\
\label{eq:thermopower}
 T\alpha_{ch}G &= \frac{1}{h}\int_0^\infty d\varepsilon\, 
\Phi(\varepsilon)
\left(\varepsilon-\mu\right)\left(-\frac{\partial f}{\partial 
\varepsilon}\right)\\
\label{eq:th_conductance}
  \frac{G_T}{T}+G\alpha_{ch}^2 &= \frac{1}{h}\int_0^\infty 
d\varepsilon\,
\Phi(\varepsilon) \left(\varepsilon-\mu\right)^2\left(-\frac{\partial 
f}{\partial
\varepsilon}\right)\,
\end{eqnarray}
where $\Phi$ is the transport function of the channel, which plays for the
channel a role equivalent to that of the density of states for the
reservoirs. $G$ is the conductance, $G_T$ the thermal conductance and 
$\alpha_{ch}$ the thermoelectric response of the channel. A simple 
interpretation of $\Phi (\varepsilon)$ is the number of channels available 
for a particle having an energy $\varepsilon$, since at zero-temperature 
its value at the Fermi energy is directly related to the conductance $G 
(T=0K) = \frac{\Phi(E_F)}{h}$.
 
For free particles of mass $M$ propagating along the 
$y$-direction and harmonically confined in the transverse ($x-z$) 
direction, it is given by~:
\begin{eqnarray}
 \Phi (\varepsilon) & = &
\sum_{n_z=0}^{\infty}\sum_{n_x=0}^{\infty}\int_0^\infty
dk_y\,\frac{\hbar k_y}{M}\mathcal{T}(k_y)
\delta\left(\varepsilon-\hbar\omega_x(n_x+1/2)-\hbar\omega_z(n_z+1/2)-\frac
{\hbar^2k_y^2}{2M}\right)\label{eq:transport_eq}\\
{} & = & \sum_{n_z=0}^{\infty}\sum_{n_x=0}^{\infty}
\mathcal{T}(\varepsilon-\hbar\omega_x(n_x+1/2)-\hbar\omega_z(n_z+1/2))
\vartheta(\varepsilon-\hbar\omega_x(n_x+1/2)-\hbar\omega_z(n_z+1/2))\,,
\end{eqnarray}
where $\mathcal{T}$ is the transmission probability which depends on the momentum or energy along the $y$ direction in the first and 
second line, respectively. The difference between the various 
transport regimes is contained in the (energy or momentum dependent) 
transmission probability $\mathcal{T}$.\\

In \eqref{eq:transport_eq}, the energy conservation condition states that
a particle entering the channel will distribute its energy between kinetic
(propagation with a certain momentum along $y$) and confinement (populating a
tranverse mode along $x$ and $z$). 
The energy dependence of $\Phi$ close to the chemical potential is 
responsible for a particle-hole asymmetry which enhances the value of the 
thermopower $\alpha_{ch}$. This effect is larger when the energy 
dependence is stronger. This is for example the case when the conduction 
regime goes from ballistic to diffusive, or when the confinement 
increases.\\

In the ballistic case, the transmission probability of the channel is equal 
to one for all energies.  For ballistic conduction, the transport function 
is in good approximation equal to $\Phi(\varepsilon) \simeq
\frac{1}{2}\left(\frac{\varepsilon}{\hbar\omega_x}+1\right)\left(\frac{
\varepsilon } {
\hbar\omega_z}+1\right)$. Since a larger confinement induces a 
steeper energy dependence of the transport function $\Phi$ around the 
chemical potential, the resulting thermopower of the channel 
$\alpha_{ch}$ is a growing function of $\omega_z$ and $\omega_x$. 

In the diffusive case, the constrain that the channel has to
obey Ohm's law leads to the following effective transmission probability
for a channel of length $\mathcal{L}$~\cite{datta1997electronic}~:
\begin{equation}\label{eq:transmission probability_general}
 \mathcal{T}(\varepsilon,\bar{V}) = 
\frac{l(\varepsilon,\bar{V})}{\mathcal{L}+l(\varepsilon,\bar{V})}\,,
\end{equation}
where $l(\epsilon,\bar{V}) = \tau_s(\bar{V},\varepsilon)v(\varepsilon)$ is 
the
energy dependent mean free path, $\tau_s$ being the scattering time, $v$
the velocity of the carriers, and the speckle power $\bar{V}$. We assume 
that the scattering time has no energy dependence, and is dominated by its 
dependence on the speckle power $\bar{V}$.

In the regime of low speckle power, where the scattering time is expected 
to be very long, the mean free path becomes typically larger than the size 
of the system $l \gg \mathcal{L}$, and the transmission probability tends 
towards one, as expected in the ballistic regime. 

In contrast, in the regime of strong speckle power, the mean free path is 
expected to be very small, leading to an effective form of the 
transmission probability
\begin{equation}
 \label{eq:effective}
 \mathcal{T}(\varepsilon,\bar{V}) \simeq \frac{\tau_s(\bar{V}) 
v(\varepsilon)}{\mathcal{L}}\,.
\end{equation}
This corresponds to the solution of the Boltzmann
equation~\cite{Ashcroft}. In this situation the resulting thermopower 
$\alpha_{ch}$ is independent of the scattering time and thus of the 
details of the speckle potential since it is the ratio of 
equations~\eqref{eq:conductance} and~\eqref{eq:thermopower}. This 
prediction is confirmed by the experimental measurements which are 
independent of the speckle power for strong disorder (see 
main text). In this diffusive regime, the timescale is given by~: 
\begin{equation}
\label{eq:TS_diffusive}
\tau_0^{-1} = 
\frac{4}{3}\frac{\nu_x\nu_y}{\nu_z}\frac{F[3/2,\bar{\mu}/k_B\bar{T}]}{F[1, 
\bar{\mu}/k_B\bar{T}]}\frac{\tau_s\sqrt{2k_B\bar{T}/M}}{\mathcal{L}}\, 
\end{equation}
where $F[n,x]$ is a Fermi-Dirac integral~\cite{AStegun}. The result 
\eqref{eq:TS_diffusive} is then used to find the behaviour of the 
scattering time $\tau_s$ as a function of speckle power with the ansatz 
$\tau_s = A(\bar{V})^{-B}$. Fitting this form to the time scale at 
strong 
speckle gives an exponent $B = 1.51 \pm 0.02$ ,$A = 6.1 \pm 
0.25\,$(a.u.) and 
$A=6.8\pm0.25\,$(a.u.) at $\nu_z=3.5$kHz and 
$\nu_z=4.95$kHz respectively. This 
form for the scattering time is then extrapolated for lower speckle powers.

\subsubsection*{Evolution of the particle imbalance and temperature
difference}

Using the linear response equations given in the main text and the
properties of the reservoirs given by 
\eqref{eq:thcoeffres_1},\eqref{eq:thcoeffres_2} and 
\eqref{eq:thcoeffres_3}, we derive equations ruling the time evolution of
the particle and temperature imbalance:  
\begin{equation}
 \tau_0\frac{d}{dt}
\left(
 \begin{array}{c}
 \Delta N/\kappa\\
  \Delta T 
 \end{array}
\right)
=
-\underline{\Lambda}
\left(
 \begin{array}{c}
 \Delta N/\kappa\\
  \Delta T 
 \end{array}
\right) , 
\underline{\Lambda}=\begin{pmatrix}
1 & -\alpha\\ 
-\frac{\alpha}{\ell} & \frac{L+\alpha^2}{\ell}
\end{pmatrix}\,,
\label{eq:Transport_equation}
\end{equation}
with $\tau_0 = \kappa/G$.
The effective thermoelectric (Seebeck) coefficient
is $\alpha\equiv\alpha_r-\alpha_{ch}$ in which the
competition between the reservoir contribution and the contribution of the channel is evident.

In the absence of any thermoelectric effect ($\alpha=0$), the time
constants for particle and thermal relaxation are $\tau_0$ and $\tau_0
L/\ell$, respectively. At low temperature (typically below $T/T_F=0.1$), 
the ratio $L/\ell$ tends to one, and the timescales for heat and particle 
transport are equal, as a consequence of the Wiedemann-Franz 
law~\cite{Ashcroft}.

The direct integration of the set of
equations~\eqref{eq:Transport_equation} provides the time evolution of the
particle imbalance and the temperature difference~:
\begin{eqnarray}
\label{eq:particle_number_result}
  (N_c-N_h)(t) &= \left\lbrace
\frac{1}{2}\left[e^{-t/\tau_-}+e^{-t/\tau_+}\right]
+\left[1-\frac{L+\alpha^2}{\ell}\right]
\frac{e^{-t/\tau_-}-e^{-t/\tau_+}}{2(\lambda_+-\lambda_-)}
\right\rbrace \Delta N_0
+\frac{\alpha\kappa}{\lambda_+-\lambda_-}\left[e^{-t/\tau_-}-e^{-t/\tau_+}
\right ]\Delta T_0\\
  (T_c-T_h)(t) &= \left\lbrace 
\frac{1}{2}\left[e^{-t/\tau_-}+e^{-t/\tau_+}\right] 
+\left[\frac{L+\alpha^2}{\ell}-1\right]\frac{e^{-t/\tau_-}-e^{-t/\tau_+}}{
2(\lambda_+-\lambda_-)}\right\rbrace \Delta
T_0+\frac{\alpha}{\ell\kappa(\lambda_+-\lambda_-)}\left[e^{-t/\tau_-}-e^{
-t/\tau_+}\right]\Delta N_0
\label{eq:temperature_result}
\end{eqnarray}
The initial temperature difference and particle imbalance are denoted by $\Delta
T_0$ and $\Delta N_0$, respectively. The inverse time-scales $\tau_\pm^{-1}=\tau_0^{-1}\lambda_\pm$ are given by
the eigenvalues of the transport matrix $\underline{\Lambda}$
\begin{equation}
 \label{eq:timescales}
 \lambda_\pm =
\frac{1}{2}\left(1+\frac{L+\alpha^2}{\ell}\right)\pm\sqrt{\frac{\alpha^2}{
\ell} +\left(\frac{1}{2}-\frac{L+\alpha^2}{2\ell}\right)^2}\,.
\end{equation} 

All the effective transport coefficients are ratios that depend only on the variable 
$\frac{\mu}{k_BT}$. 

\subsubsection*{Fitting procedure}

\begin{figure}[htb]
\def \thefigure{S3}
\begin{subfigure}[]
                \centering
\includegraphics[width=0.45\textwidth]{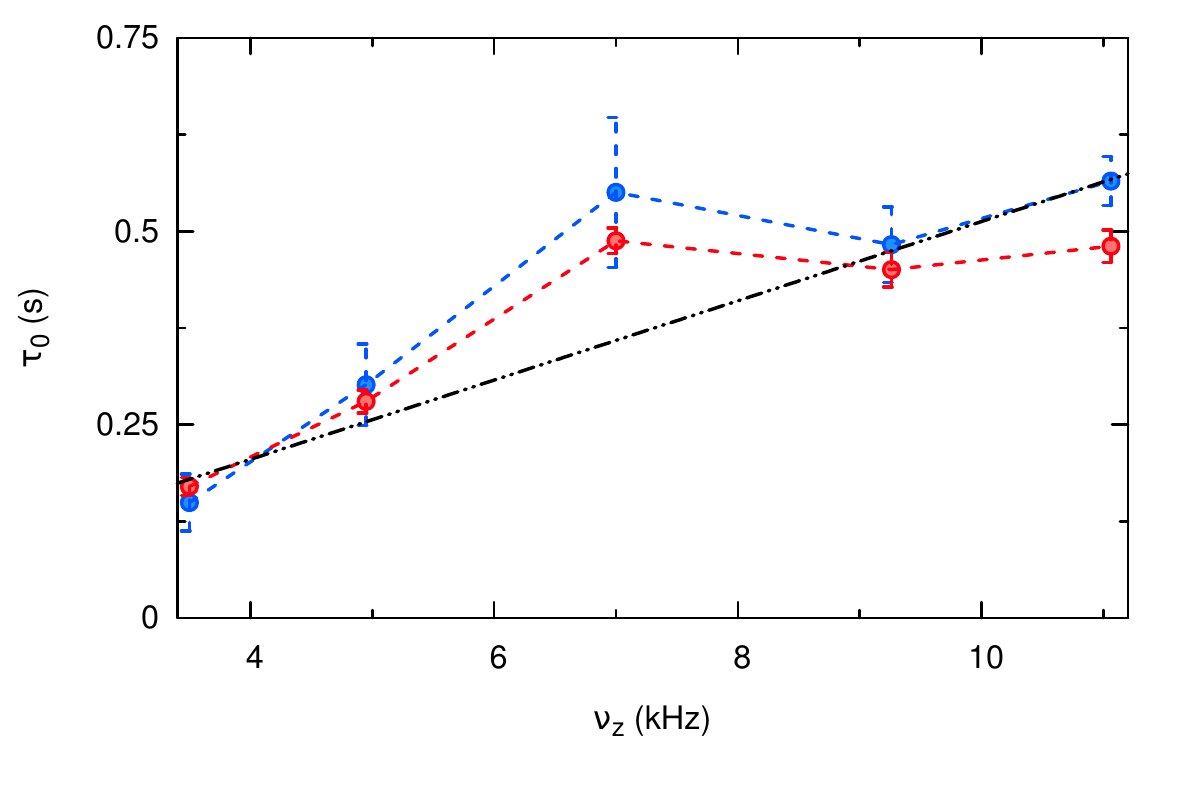}
                \label{fig:TSB}
\end{subfigure}
\begin{subfigure}[]
                \centering
\includegraphics[width=0.45\textwidth]{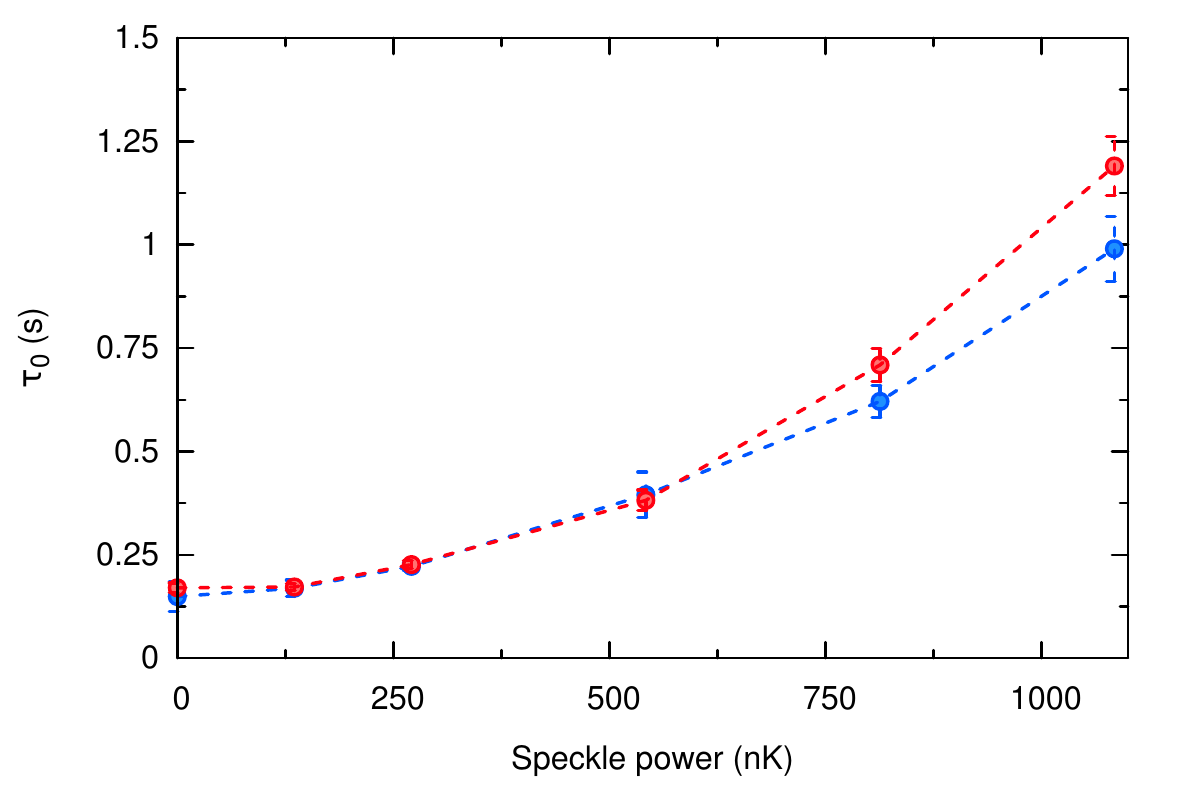}
                \label{fig:TSD}
\end{subfigure}
\caption{Timescale $\tau_0$ extracted from a fit of the particle number imbalance 
in sequence (1) (in blue) and sequence (2) (in red), vs confinement in a 
ballistic channel (left panel), and vs speckle power at fixed confinement 
$\nu_z = 3.5\,$kHz (right panel). The black line in the left panel 
corresponds to an ab initio prediciton for the timescale from the 
expressions~\eqref{eq:thcoeffres_1} and~\eqref{eq:conductance}.}
\label{fig:TS}
\end{figure}

We have used the solution~\eqref{eq:particle_number_result} and
\eqref{eq:temperature_result} to the transport equations in order to 
extract the time scale $\tau_0$ from the experimental data.
For each value of the parameters (confinement along the $z$-direction,
disorder strength and temperature), two experimental sequences have been 
performed :
\begin{enumerate}[(1)]
 \item A 'thermoelectric' sequence, with dominating initial temperature 
difference which is prepared by the procedure described in the main text. 
The resulting evolution is a coupled evolution of the particle and 
temperature differences as described
in the main text.
 \item A 'decay' sequence, with initial particle imbalance, at constant 
temperature equal to the final temperature in sequence (1).
\end{enumerate}
The time scale shown in the main text is extracted from the first 
sequence. However, in order to check consistency we compare in 
fig.~\ref{fig:TS} the time scales extracted from the 
two sequences. The two independent fits agree very well also in the 
ballistic case with the parameter free theoretical calculations. 

\subsubsection*{Efficiency}
In this section we will derive the expression for the efficiency employed 
in the main text. In order to do this, we use the 
conservation of the total energy and particle number present in the setup, i.e.
\begin{eqnarray}
 dE_{tot}=d(E_{\mathrm{ch}}+E_c+E_h)&=0\\
 dN_{tot}=d(N_c+N_h)&=0\,,
\end{eqnarray}
where the subscript $\mathrm{ch}$ denotes quantities in the channel. For the particle conservation, we have assumed that in the channel the particles do not accumulate, i.e. $dN_{\mathrm{ch}}=0$.
Reexpressing the change of the energy by the change of entropy and particle 
number in each part of the system $ dE_x  = T_xdS_x+\mu_xdN_x$ and using 
particle number conservation, we obtain
\begin{equation}
dE_{tot}=  \left(\mu_h-\mu_c\right)dN_h +\left(T_{\mathrm{ch}}dS_{\mathrm{ch}}+T_cdS_c+T_hdS_h\right)=0.
\end{equation}
 Reordering and the derivation with respect to time leads to an expression 
for the total change of entropy per time given by
\begin{equation}
 \frac{d(S_{\mathrm{ch}}+S_h+S_c)}{dt} = -\frac{\Delta T}{2\bar{T}}(\dot{S}_h 
  - \dot{S}_c)-\frac{\Delta \mu}{2\bar{T}}(\dot{N}_h  
 - \dot{N}_c) = -\frac{\Delta T}{2\bar{T}}I_S-\frac{\Delta 
\mu}{2\bar{T}}I_N\,.
\end{equation} 
Here we used the average temperature $\bar{T}=\frac 1 2 (T_c+T_h)=T_{\mathrm{ch}}$.

Thus, the efficiency defined by the ratio of the work to irreversible heat
\begin{equation}
 \label{eq:efficiency_result}
 \eta = -\frac{\int (\dot{N_c}-\dot{N_h})\cdot(\mu_c-\mu_h)dt}{\int 
(\dot{S_c}-\dot{S_h})\cdot(T_c-T_h)dt}
\end{equation}
measures the efficiency relative to a reversible process : $\eta=1$ when 
the evolution is reversible. 

Using the solution for particle 
imbalance~\eqref{eq:particle_number_result} and temperature 
difference~\eqref{eq:temperature_result} yields the following expression as 
a function of the effective transport coefficients~:
\begin{equation}
 \eta = \frac{-\alpha\alpha_r}{\ell+L+\alpha^2-\alpha\alpha_r}\,.
\end{equation}


\end{document}